\pdfoutput=1
\documentclass{article} 
\usepackage{preprint,times}

\usepackage{amsmath,amsfonts,bm}

\def\eqref#1{equation~\ref{#1}}

\def\plaineqref#1{\ref{#1}}

\def\1{\bm{1}}

\DeclareMathAlphabet{\mathsfit}{\encodingdefault}{\sfdefault}{m}{sl}
\SetMathAlphabet{\mathsfit}{bold}{\encodingdefault}{\sfdefault}{bx}{n}

\usepackage{hyperref}
\usepackage{url}
\usepackage{algorithm}
\usepackage{algorithmic}
\usepackage{graphicx}
\usepackage{amsthm}
\usepackage{booktabs}
\usepackage{enumitem}
\usepackage{placeins}

\newtheorem{theorem}{Theorem}
\newtheorem{proposition}[theorem]{Proposition}

\title{OPINE: Orthogonal Projection in Invertible Nonlinear Embeddings for Turbulent Flows}

\author{Anqiao Ouyang\\
Faculty of Mathematics\\
University of Waterloo\\
Waterloo, ON N2L 3G1, Canada\\
Department of Aerospace Engineering\\
San Diego State University\\
San Diego, CA 92139, USA \\
\And
Hongyi Ke\\
Department of Aerospace Engineering\\
San Diego State University\\
San Diego, CA 92139, USA \\
\And
Qi Wang\\
Department of Aerospace Engineering\\
San Diego State University\\
San Diego, CA 92139, USA \\
\texttt{qwang4@sdsu.edu} \\
}

\begin{document}

\maketitle

\begin{abstract}
Proper orthogonal decomposition (POD) represents physical fields with orthogonal modes ordered by variance, so that one decomposition gives reduced coordinates in every dimension, but it needs many modes when the data lie near a curved manifold. Nonlinear autoencoders are more compact, but their information loss happens inside a nonlinear encoder, and a model trained with one latent dimension gives no model with a smaller one. We introduce Orthogonal Projection in Invertible Nonlinear Embeddings (OPINE), which learns an invertible nonlinear change of coordinates and performs POD in the transformed coordinates. Because the map is invertible, all information loss occurs in the POD projection, and ordering its basis after training gives reduced coordinates of every smaller dimension from a single model. Training alternates gradient updates of the map with a damped update that tracks the evolving POD subspace, and we prove that this update converges locally to the exact POD subspace. We evaluate OPINE on synthetic manifolds, chaotic Kuramoto--Sivashinsky dynamics, forced Burgers turbulence, two-dimensional Kolmogorov flow, and three-dimensional minimal channel flow. OPINE is more accurate than POD at every tested retained dimension above one, and on the Kuramoto--Sivashinsky and Burgers dynamics, it is more accurate than convolutional and fully connected autoencoders by more than two orders of magnitude at the largest retained dimension.
\end{abstract}

\section{Introduction}

A snapshot of a turbulent flow has thousands to millions of degrees of
freedom, yet its dominant dynamics often lie near a much lower-dimensional
set. Reduced coordinates that describe this set are the basis of data
compression, flow analysis, and reduced-order models. The standard method for
obtaining them is proper orthogonal decomposition (POD), known in statistics
as principal component analysis. POD modes are orthogonal and ordered by the
variance they capture, and for every \(r\) the leading \(r\) modes span the
best \(r\)-dimensional linear subspace
\citep{lumley1970stochastic,berkooz1993proper,taira2017modal}. One
decomposition therefore provides reduced models of every size, and each
coordinate has a clear physical meaning.

This optimality, however, holds only among linear subspaces. Curved and
transport-dominated solution sets, such as traveling waves, shocks, and the
chaotic attractors of turbulent flows, can need many modes even when they have low-dimensional nonlinear parametrizations \citep{greif2019decay,lee2020model}.
Autoencoders overcome this limitation with nonlinear encoders and decoders,
and they have improved on POD for fluid fields
\citep{kramer1991nonlinear,milano2002neural,murata2020nonlinear}. However, a standard autoencoder has
neither orthogonal modes nor an ordering, its encoder can map different states
to the same latent vector, and a model trained with one latent dimension gives
no model with a smaller one. Nested dropout, Matryoshka
representations, and hierarchical autoencoders order the latent variables
\citep{rippel2014nested,kusupati2022matryoshka,fukami2020hierarchical}, but
the truncation still happens inside a nonlinear encoder.

\begin{figure}[!ht]
 \centering
 \includegraphics[width=\linewidth]{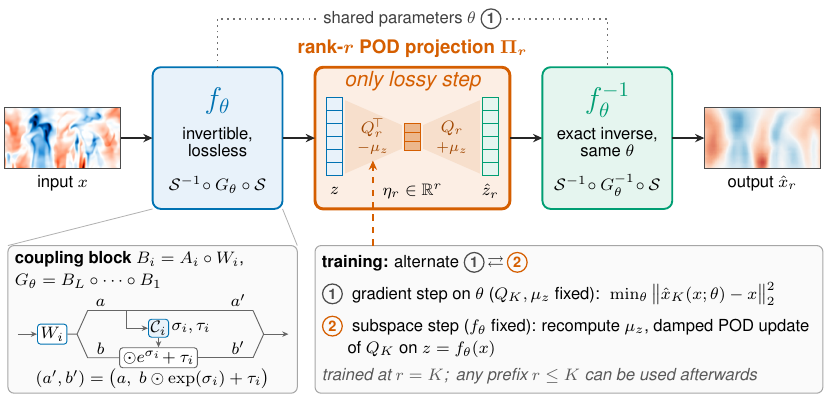}
 \caption{
 OPINE for a scalar field: only the rank-$r$ POD projection of
 $z=\operatorname{vec}f_\theta(x)$ loses information, and $f_\theta$ and its
 exact inverse share $\theta$. Training alternates the two steps at bottom
 right (Section~\ref{sec:training}). For multi-component fields,
 $\mathcal S$ is omitted and $f_\theta=G_\theta$.
 }
 \label{fig:opine_architecture}
\end{figure}

Our key idea is to keep POD as the \emph{only} bottleneck, but to apply it in
transformed coordinates. In \emph{Orthogonal Projection in Invertible Nonlinear
Embeddings} (OPINE), an invertible map, parametrized by a neural network, transforms the field
without reducing its dimension, POD removes dimensions in the transformed
coordinates, and the inverse map returns the result to physical space
(Figure~\ref{fig:opine_architecture}).
Because the map is invertible and keeps the dimension of the field, no
information is lost before the projection, and the whole model reduces to the
identity when all coordinates are kept. OPINE can therefore be viewed as a
search for an observable of the field, that is, a nonlinear change of
variables in which the data lie close to a linear subspace, similar to the
feature map of kernel PCA \citep{scholkopf1998kpca} but invertible, so that no
pre-image problem arises \citep{mika1999kernel}. Once this
observable is found, the properties of POD, including orthogonal, ordered,
and nested modes, come at no additional cost. Training is a coupled problem, because the POD basis changes whenever the
map changes, and a sudden change of the basis would also change the loss suddenly. We therefore alternate gradient updates of the map with a damped,
aligned step of orthogonal iteration. Our contributions are:
\begin{itemize}[leftmargin=*]
 \item \textbf{Method:} a factorization of nonlinear dimensionality reduction into an
 invertible change of coordinates and an explicit POD projection, which
 confines all information loss to the projection and gives reduced coordinates of every smaller dimension from one model (Section~\ref{sec:method});
 \item \textbf{Training and theory:} a scheme that tracks the POD subspace without
 differentiating through an eigendecomposition, together with a proof that
 the damped update converges to the exact POD subspace and follows it
 closely while the map changes (Section~\ref{sec:training};
 Appendix~\ref{app:pod});
 \item \textbf{Experiments:} from synthetic manifolds and one-dimensional chaotic
 dynamics to two- and three-dimensional turbulence, in which OPINE improves
 on POD at every retained dimension above one, and ablations that support the choice of the POD projection
 (Section~\ref{sec:experiments}).
\end{itemize}

\section{Related Work}

OPINE is related to three lines of work: autoencoders and nonlinear model
reduction, ordered and invertible representations, and subspace tracking.

\paragraph{Autoencoders and nonlinear model reduction.}
Autoencoders learn a nonlinear encoder and decoder through a low-dimensional
bottleneck and are the standard nonlinear generalization of POD
\citep{kramer1991nonlinear,hinton2006reducing}; variational autoencoders add a
probabilistic latent space \citep{kingma2014vae}. In fluid mechanics,
autoencoders have compressed near-wall turbulence \citep{milano2002neural},
decomposed flows into nonlinear modes \citep{murata2020nonlinear}, built
hierarchical or nearly orthogonal latent modes
\citep{fukami2020hierarchical,eivazi2022towards}, and learned coordinates for
Koopman and inertial-manifold models
\citep{lusch2018deep,champion2019data,linot2020deep}. In reduced-order
modeling, convolutional autoencoders provide nonlinear trial manifolds for
advection-dominated systems \citep{lee2020model,maulik2021reduced}, and POD
precompression makes them cheaper to train \citep{fresca2022poddrom}. Classical alternatives transport the data before a
linear projection, as in shifted POD and registration
\citep{reiss2018shifted,taddei2020registration}, or add a polynomial
correction to a POD subspace, as in quadratic manifolds
\citep{geelen2023quadratic}; these fix the form of the transformation. OPINE
instead learns a general invertible map and applies POD \emph{after} it, so
that POD remains the only bottleneck.

\paragraph{Ordered and invertible representations.}
Invertible networks such as Real NVP, Glow, and invertible residual networks
provide flexible bijective maps with tractable inverses
\citep{dinh2017realnvp,kingma2018glow,behrmann2019iresnet}, and injective or
manifold flows use such maps to learn low-dimensional manifolds
\citep{brehmer2020flows,caterini2021rectangular,kothari2021trumpets,cunningham2022pcf}.
Closest to OPINE are nested-dropout flows, which order the coordinates of an invertible network and reconstruct by zeroing its trailing coordinates
\citep{rippel2014nested,bekasov2020ordering}, and Neural-PCA, which couples a
flow to a PCA layer for likelihood-based representation learning
\citep{li2022neural}. In principle, an unrestricted invertible map could
absorb a global rotation, but stacks of local coupling layers cannot represent
one cheaply. OPINE therefore keeps its bottleneck close to the POD subspace of
the transformed data and updates it in closed form, which we show matters in
practice. It is also trained for field reconstruction rather than likelihood.

\paragraph{Subspace tracking.}
Online PCA and subspace-tracking algorithms update dominant eigenspaces as
the data evolve \citep{oja1982simplified,yang1995projection}, whereas
backpropagating through an eigendecomposition
\citep{ionescu2015matrix,wang2019backpropagation} is fragile near the
repeated eigenvalues of paired Fourier modes. We use damped, aligned
orthogonal iteration.


\section{Method}
\label{sec:method}

Sections~\ref{sec:representation}--\ref{sec:pod} present our first
contribution, the factorization, and Section~\ref{sec:training} the second,
the training scheme. Let \(x\in\mathbb{R}^{n_c\times N_1\times\cdots\times N_d}\) be a
discretized field with \(n_c\) components on a grid of
\(N_1\times\cdots\times N_d\) points, and thus \(N=n_c\prod_jN_j\) degrees of
freedom. OPINE places an explicit projection inside an invertible change of
coordinates:
\begin{equation}
 x\xrightarrow{f_\theta}z\xrightarrow{\Pi_r}\hat z_r
 \xrightarrow{f_\theta^{-1}}\hat x_r .
 \label{eq:opine_pipeline}
\end{equation}
Because \(f_\theta\) is bijective, the projection \(\Pi_r\) is the only step
that loses information.

\subsection{Representation}
\label{sec:representation}

We first define the reduced coordinates. Let
\(z=\operatorname{vec}f_\theta(x)\in\mathbb{R}^N\), let \(\mu_z\) be the
mean of the transformed data, and let \(Q_K=[q_1,\ldots,q_K]\) be an
orthonormal basis of the learned \(K\)-dimensional subspace. For \(r\leq K\)
and \(Q_r=[q_1,\ldots,q_r]\), the reduced coordinates and the reconstruction
are
\begin{equation}
 \eta_r=Q_r^\top(z-\mu_z)\in\mathbb{R}^r,
 \qquad
 \hat z_r=\mu_z+Q_r\eta_r,
 \qquad
 \hat x_r=f_\theta^{-1}\bigl(\operatorname{unvec}\hat z_r\bigr).
 \label{eq:coefficients}
\end{equation}
Here \(\Pi_r:z\mapsto\hat z_r\), and \(r\) counts real coefficients. The
retained dimension \(D\) of the experiments equals \(r\), and each model is
trained with \(K=D\). When all transformed coordinates are kept (\(r=K=N\)), the
reconstruction is exact. Because \(\Pi_r\) is a projection and \(f_\theta\)
is bijective, \(x\mapsto\hat x_r\) is a nonlinear projection onto an
\(r\)-dimensional manifold, the kind of trial manifold used by manifold
Galerkin and least-squares Petrov--Galerkin models \citep{lee2020model}.

\subsection{Invertible Map}
\label{sec:coupling}

The representation above holds for any invertible \(f_\theta\); we now
describe the one we use. It is a stack
\(G_\theta=B_L\circ\cdots\circ B_1\) of blocks \(B_i=A_i\circ W_i\), where
\(W_i\) is an invertible channel mixing \citep{kingma2018glow} and \(A_i\) is
an affine coupling layer \citep{dinh2017realnvp} that splits the channels
into \((a,b)\) and maps
\begin{equation}
 (a,b)\mapsto\bigl(a,\;b\odot\exp(\sigma_i)+\tau_i\bigr),
 \qquad (\sigma_i,\tau_i)=\mathcal C_i(a).
 \label{eq:coupling}
\end{equation}
This layer can be inverted in closed form for any conditioner network
\(\mathcal C_i\). Multi-component fields use their physical components as
channels, so that \(f_\theta=G_\theta\). Scalar fields first pass through an
invertible space-to-depth rearrangement \(\mathcal S\), which folds small
blocks of neighboring grid points into channels, so that
\(f_\theta=\mathcal S^{-1}\circ G_\theta\circ\mathcal S\). We also tried
invertible residual networks and monotone piecewise-quadratic maps, but affine
coupling layers were the most accurate and the fastest to invert
(Appendix~\ref{app:details}).

\subsection{POD Bottleneck}
\label{sec:pod}

The bottleneck is the POD of the transformed data. For fixed \(\theta\) and
transformed training states \(z_i\) with mean \(\mu_z\) and covariance
\(C_z\), the rank-\(K\) POD basis
\begin{equation}
 Q_K^\star\in\arg\min_{Q^\top Q=I_K}\sum_i\bigl\|(I-QQ^\top)(z_i-\mu_z)\bigr\|_2^2
 \label{eq:pod}
\end{equation}
spans the leading \(K\)-dimensional eigenspace of \(C_z\), whose eigenvalues
we denote \(\lambda_1\geq\cdots\geq\lambda_N\). 

\subsection{Training}
\label{sec:training}

It remains to train the map and the basis together. The map \(f_\theta\)
minimizes the reconstruction error, while \(Q_K\) and \(\mu_z\) must follow
the POD of the transformed data, which changes as \(\theta\) changes. We keep
\(Q_K\) tied to that POD, rather than learning it as a free parameter, so that the retained subspace can be ordered by variance. Training alternates two updates, neither of which differentiates
through an eigendecomposition. First, with \(Q_K\) and \(\mu_z\) fixed, we
take a round of gradient steps on
\begin{equation}
 \mathcal L_{\mathcal B}(\theta)=\frac{1}{|\mathcal B|}\sum_{x\in\mathcal B}
 \bigl\|\hat x_K(x;\theta)-x\bigr\|_2^2
 \label{eq:loss}
\end{equation}
over minibatches \(\mathcal B\), with \(\hat x_K\) from
Eq.~\plaineqref{eq:coefficients}. Second, with \(f_\theta\) fixed, we
recompute \(\mu_z\) and take one damped, aligned step of orthogonal
iteration:
\begin{equation}
 \widetilde Q_K=\operatorname{qf}(C_zQ_K),
 \qquad
 R=\arg\min_{R^\top R=I_K}\bigl\|\widetilde Q_KR-Q_K\bigr\|_F=U_RV_R^\top,
 \label{eq:power}
\end{equation}
\begin{equation}
 Q_K\leftarrow\operatorname{qf}\bigl[(1-\alpha)Q_K+\alpha\widetilde Q_KR\bigr],
 \label{eq:q_update}
\end{equation}
where \(\operatorname{qf}\) is the orthonormal factor of a thin QR
factorization and \(\widetilde Q_K^\top Q_K=U_R\Sigma V_R^\top\) is an SVD.
Orthogonal iteration moves the subspace toward the dominant eigenspace but
may rotate the basis within it, and the Procrustes rotation \(R\) undoes this
rotation so that the old and new bases can be interpolated. Since
\(Q_K^\top\widetilde Q_KR=V_R\Sigma V_R^\top\succeq0\), the update is well
defined for every \(\alpha\in[0,1]\), and a smaller \(\alpha\) gives stronger
damping. Appendix~\ref{app:pod} proves that this update converges
to the exact POD subspace and follows it closely while \(f_\theta\) changes.

After training, we diagonalize \(Q_K^\top C_zQ_K=E\Lambda_QE^\top\) and set
\(Q_K\leftarrow Q_KE\). This leaves the projection unchanged and orders the
basis by variance, so each prefix \(Q_r\) spans the most energetic
\(r\)-dimensional subspace of \(\operatorname{span}(Q_K)\), and one model gives reduced coordinates for every \(r\leq K\)
(Section~\ref{sec:nested}).

\section{Experiments}
\label{sec:experiments}

Our third contribution is an evaluation of OPINE on datasets of increasing
dimension, from synthetic one-dimensional profiles to three-dimensional
turbulence. We ask four questions:
whether OPINE finds transformed coordinates in which data of known low
dimension have low rank (\textbf{Q1}; Section~\ref{sec:synthetic}); whether it improves on POD and
autoencoders with the same number of retained coefficients (\textbf{Q2};
Sections~\ref{sec:onedim}--\ref{sec:minimal3d}); whether its improvement over POD comes from the transformed coordinates or
from the choice of projection (\textbf{Q3}; Section~\ref{sec:synthetic}); and whether one trained model is accurate at every
smaller \(D\) (\textbf{Q4}; Section~\ref{sec:nested}).

\subsection{Baselines and Evaluation}
\label{sec:baselines}

To keep the comparison fair, every method stores the same number \(D\) of
real coefficients per snapshot: \(r=D\) for POD and OPINE, and a
\(D\)-dimensional latent vector for the autoencoders. POD is Eq.~\plaineqref{eq:coefficients} with \(f_\theta\) the identity. The CNN autoencoder compresses the field to
its latent vector with strided convolutions and boundary-aware padding, and
reconstructs it with a mirrored decoder (Figure~\ref{fig:cnn_baseline} in
Appendix~\ref{app:details}). For the one-dimensional problems we also report
a fully connected (MLP) autoencoder. OPINE also has far fewer parameters: for the turbulent flows, it has \(6\times10^4\) to \(10^5\) parameters, whereas the
CNN autoencoders have \(10^6\) to \(2\times10^7\). All learned models are
trained separately at each \(D\) and selected on validation data
(Appendix~\ref{app:details}). We report the test normalized mean squared
error (NMSE)
\(\sum_{i\in\mathcal T}\|x_i-\hat x_i\|_2^2/\sum_{i\in\mathcal T}\|x_i\|_2^2\)
on a held-out set \(\mathcal T\) shared by all methods.

\subsection{Synthetic Manifolds and Ablations}
\label{sec:synthetic}

We begin with synthetic data whose low-dimensional parametrization is known exactly. The family \(x_\phi(s)=\tanh[\sin(s+\phi)]\) on a periodic grid, with
a random phase \(\phi\sim\mathcal U(0,2\pi)\), is a one-parameter set of
traveling profiles. It is not low rank in the original coordinates, because
the \(\tanh\) generates all odd harmonics, but it is exactly rank~\(2\) in the
invertible coordinates
\(\operatorname{atanh}x_\phi=\sin s\cos\phi+\cos s\sin\phi\). We train at
\(D=2\) with \(32\) samples and ten seeds. OPINE finds such coordinates (Q1):
two POD modes of its transformed coordinates hold \(99.99\%\) of the transformed
variance, and its test error is lower than that of POD on every seed and
\(28\) times lower on average, whereas the CNN autoencoder overfits the small
training set.

\begin{figure}[!ht]
 \centering
 \includegraphics[width=\linewidth]{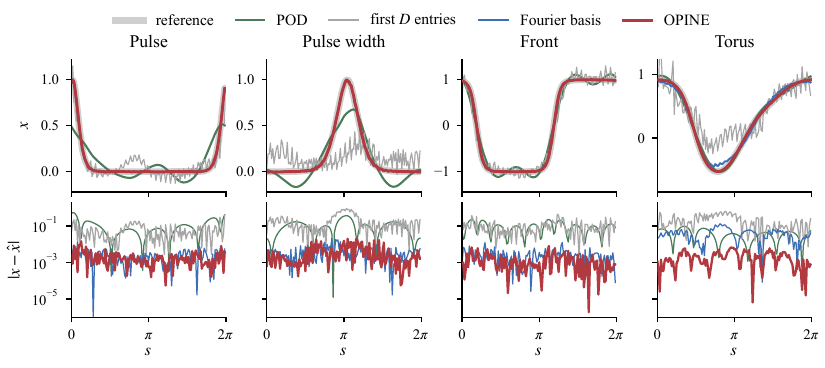}
 \caption{
    Sample reconstructions for the bottleneck ablation of
    Table~\ref{tab:onedim} at $D=4$. Top: a test profile (grey) and its
    reconstructions; the profile is the test sample with the median OPINE
    error. Bottom: pointwise error on a logarithmic scale. POD misses the
    shape of each profile, and keeping the first $D$ entries of $z$ gives noisy
    profiles. The error of OPINE is about two orders of magnitude below that of
    POD, and on Front and Torus it is also well below that of the fixed Fourier
    basis.
    }
    \label{fig:bneck_recon}
\end{figure}

We then use four synthetic families of profiles on \(128\) points of
\(s\in[0,2\pi)\) to identify the source of the improvement over POD (Q3). In
\emph{Pulse}, \(x=\operatorname{sech}^2(d/0.3)\), where \(d\) is the periodic
distance from a random center \(c\); its states form a closed loop.
\emph{Pulse width} is the same pulse with a random width in \([0.2,0.6]\)
instead of \(0.3\). In \emph{Front}, \(x=\tanh[\kappa\sin(s-c)]\) with a
random position \(c\) and steepness \(\kappa\in[1,5]\). In \emph{Torus},
\(x=\tanh[\sin(s+\phi_1)+\sigma(\cos(2s+\phi_2))]\) with two random phases and
the logistic function \(\sigma\); its states form a torus. POD needs many
modes for all four, because the parameters move or deform the profiles.
Table~\ref{tab:onedim} retrains the same invertible map with different projections and otherwise identical training. Keeping the first \(D\) entries of \(z\), as nested-dropout flows do
at a fixed dimension, is usually no more accurate than plain POD. A fixed Fourier basis improves on
POD on three of the four families, but OPINE is the most accurate on all
four. In Figure~\ref{fig:bneck_recon}, the pointwise error of OPINE is about
two orders of magnitude below that of POD along the whole profile, whereas
keeping the first \(D\) entries of \(z\) gives noisy reconstructions. For these translation-invariant
families, POD of the raw data is itself a Fourier basis ordered by variance.
The fixed Fourier basis differs from OPINE because the invertible map is not
translation-equivariant, so the POD modes of \(z\) are not Fourier modes, and
because the basis is ordered by wavenumber rather than by variance. Table~\ref{tab:onedim} is an ablation of the projection with the
architecture and training of the invertible map held fixed; OPINE is compared with the
autoencoders on the dynamical systems below, where the benefit of the
transformed coordinates is largest.

On the same families, damped tracking is as accurate as exact recomputation
of the POD, consistent with Appendix~\ref{app:pod}, whereas freezing the subspace
at its initial POD roughly triples the error on Pulse. Tracking is also much cheaper than differentiating through an eigendecomposition
(Appendix~\ref{app:synthetic_results}).

\begin{table}[!ht]
 \centering
 \footnotesize
 \caption{Bottleneck ablation at $D=4$: median test NMSE. POD projects the
 raw data. The other three columns retrain the same invertible map with a
different projection: keeping the first $D$ entries of $z$ and setting the others to their training mean, onto a fixed
Fourier basis, or onto the POD subspace of the transformed field, which is
OPINE. Best in bold. Figure~\ref{fig:bneck_recon} shows sample
 reconstructions.}
 \label{tab:onedim}
 \vspace{3pt}
 \setlength{\tabcolsep}{5pt}
 \begin{tabular}{lcccc}
 \toprule
 & & \multicolumn{3}{c}{same invertible map, projection onto:} \\
 \cmidrule(l){3-5}
 & POD & first $D$ entries & Fourier basis & OPINE \\
 \midrule
Torus & $4.8{\times}10^{-3}$ & $5.3{\times}10^{-1}$ & $7.1{\times}10^{-3}$ & $\mathbf{1.9{\times}10^{-5}}$ \\
Front & $1.0{\times}10^{-2}$ & $2.4{\times}10^{-1}$ & $3.9{\times}10^{-5}$ & $\mathbf{6.0{\times}10^{-6}}$ \\
Pulse & $3.8{\times}10^{-1}$ & $3.4{\times}10^{-1}$ & $4.1{\times}10^{-4}$ & $\mathbf{1.3{\times}10^{-4}}$ \\
Pulse width & $2.5{\times}10^{-1}$ & $5.6{\times}10^{-1}$ & $9.6{\times}10^{-4}$ & $\mathbf{6.8{\times}10^{-4}}$ \\
 \bottomrule
 \end{tabular}
\end{table}

\begin{figure}[!ht]
 \centering
 \includegraphics[width=\linewidth]{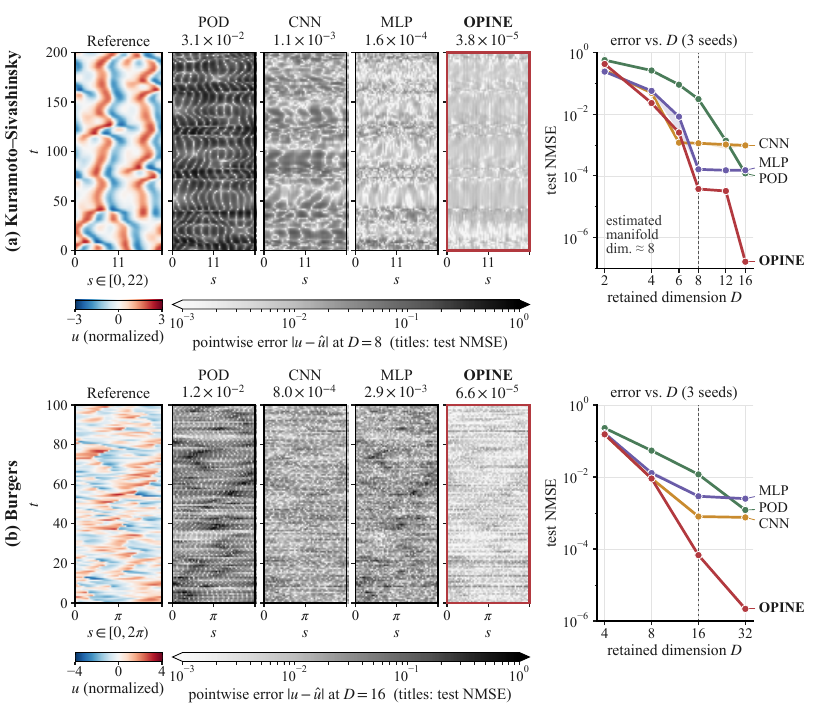}
 \caption{
 Kuramoto--Sivashinsky (a) and forced Burgers turbulence (b). Left: a test
 trajectory in space and time. Middle: pointwise error of each method at
 the $D$ marked by the dashed line on the right, on a shared logarithmic
 scale (darker is worse); titles give the test NMSE of the run shown. Right: test NMSE
 versus $D$ (median and range over three seeds).
 }
 \label{fig:onedim}
\end{figure}
\subsection{One-Dimensional Dynamics}
\label{sec:onedim}

Having confirmed that OPINE finds known low-dimensional coordinates, we turn to two
one-dimensional systems whose dynamics concentrate near a low-dimensional set
that is not known in advance. The Kuramoto--Sivashinsky (KS) equation
\(u_t+uu_s+u_{ss}+u_{ssss}=0\) on a periodic domain of length \(22\) models
flame fronts and thin-film flows, and its chaotic attractor lies close to an
inertial manifold of dimension about \(8\) \citep{linot2020deep}. Forced
Burgers turbulence, \(u_t+uu_s=\nu u_{ss}+F\), is driven by random forcing of
the largest scales that is uncorrelated in time. It produces a statistically
stationary field of steep, moving shocks, a standard transport-dominated test case for linear reduction. Each system has eight independent trajectories,
which are split by trajectory (Appendix~\ref{app:synthetic_results}).

Figure~\ref{fig:onedim} summarizes both systems. The left panels show a test
trajectory in space and time: KS forms a slowly drifting cellular pattern,
and Burgers a field of sharp fronts. The middle panels show where each method
errs at a fixed \(D\), with darker regions marking larger pointwise error. POD has large errors everywhere, the autoencoders have errors concentrated
along the cells and shocks, and OPINE's error is small everywhere. The right panels show the
test NMSE over \(D\). From \(D=8\) onward, OPINE is at least as accurate as every other method, and its error keeps falling where the autoencoders level off, so that
at the largest \(D\) it is more than two orders of magnitude below them. On
KS, its error drops sharply near \(D=8\), the manifold dimension estimated by
\citet{linot2020deep}.

\subsection{Two-Dimensional Kolmogorov Flow}
\label{sec:kolmogorov}

We now move from one to two space dimensions. Kolmogorov flow is
two-dimensional Navier--Stokes flow driven by a sinusoidal body force,
\(\partial_t\omega+\mathbf u\cdot\nabla\omega=\mathrm{Re}^{-1}\nabla^2\omega-k_f\cos(k_fy)\),
on a doubly periodic domain. At \(\mathrm{Re}=100\) with forcing wavenumber
\(k_f=4\), the flow is chaotic, with coherent vortices that form, merge, and
decay. We reduce vorticity snapshots on a \(128\times128\) grid from \(100\)
independent trajectories, split by trajectory so that test snapshots are
never close in time to training snapshots (Appendix~\ref{app:details}).

\begin{figure}[!ht]
 \centering
 \includegraphics[width=0.78\linewidth]{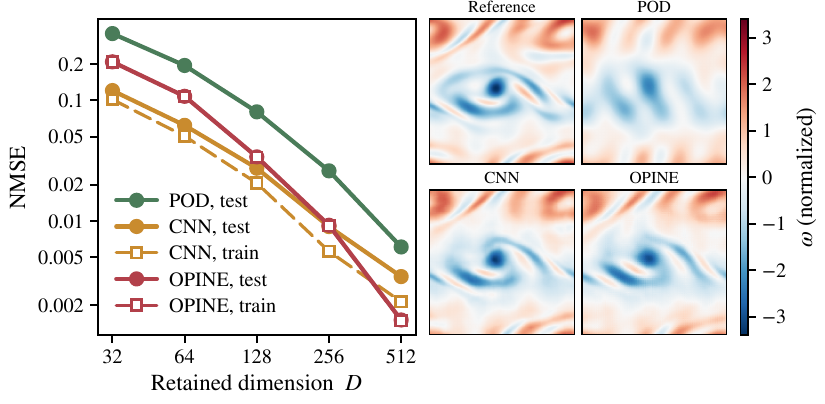}
 \caption{
 Kolmogorov flow. Left: test NMSE versus $D$ (and train NMSE of the learned
 models). Right: vorticity of a test snapshot and its reconstructions at
 $D=64$; POD blurs the vortex cores and filaments that the CNN and OPINE
 retain.
 }
 \label{fig:kolmogorov_comparison}
\end{figure}

\begin{figure}[!ht]
 \centering
 \includegraphics[width=\linewidth]{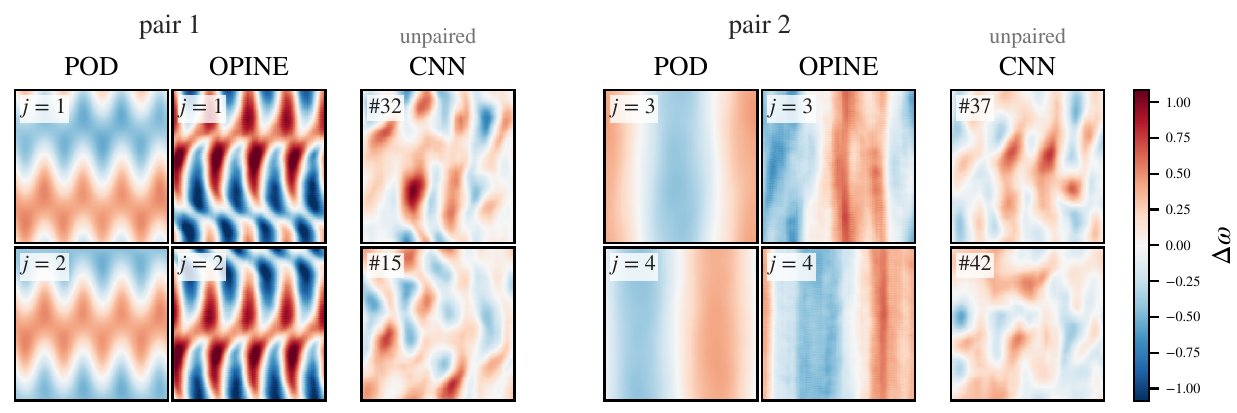}
 \caption{
 Decoded reduced coordinates for Kolmogorov flow at $D=64$, each
 activated at one training standard deviation. POD and OPINE directions
 form phase-shifted pairs ($j$: index in variance order), and OPINE's are
 sharper. CNN latent coordinates have no ordering or pairing, so for each
 pair we show the two most correlated ones (\#: latent index).
 }
 \label{fig:kolmogorov_latent_directions}
\end{figure}

Figure~\ref{fig:kolmogorov_comparison} shows that OPINE improves on POD at
every \(D\), by up to \(75\%\), and that the margin grows with \(D\) (Q2). The
CNN autoencoder, which has \(20\) to \(300\) times more parameters, is more
accurate at small \(D\), and OPINE is more accurate at the largest \(D\). The reconstructions on the right show the source of the larger POD error:
with the same number of coefficients, POD blurs the vortex cores and the thin filaments between
them, which the nonlinear methods retain. Figure~\ref{fig:kolmogorov_latent_directions}
decodes single reduced coordinates back to vorticity. The POD and OPINE coordinates come in phase-shifted pairs of the same
spatial scale, as expected for a statistically homogeneous direction, and
OPINE's are sharper; the CNN latent coordinates are not paired.

\begin{figure}[!ht]
 \centering
 \includegraphics[width=\linewidth]{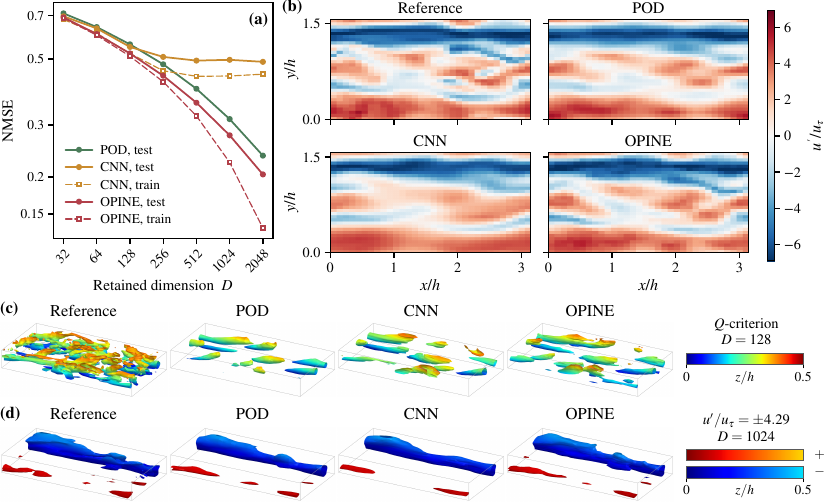}
 \caption{
 Three-dimensional minimal channel. (a) Test NMSE versus retained dimension
 $D$, with the train NMSE of the CNN and OPINE dashed. (b) Streamwise
 fluctuation in a wall-parallel plane of a test snapshot and its
 reconstructions at $D=1024$. (c) $Q$-criterion isosurfaces at a common
 isovalue for $D=128$ in another test snapshot, and (d)
 streamwise-fluctuation isosurfaces $u'/u_\tau=\pm4.29$ ($u_\tau$: friction
 velocity) at $D=1024$ in the snapshot of (b); color indicates
 wall-normal position.
 }
 \label{fig:minimal3d}
\end{figure}

\subsection{Three-Dimensional Minimal Channel Flow}
\label{sec:minimal3d}

Finally, we turn to wall-bounded turbulence, which is organized around
elongated streaks of slow and fast fluid and the vortices that regenerate them
near the wall. We reduce full three-dimensional velocity fields in a minimal
channel at friction Reynolds number \(Re_\tau=180\), as in the direct
numerical simulation of \citet{kim1987turbulence}. This small periodic box, in the spirit of the
minimal flow unit of \citet{jimenez1991minimal}, holds only a few near-wall
streaks and the vortices that sustain them. Each snapshot has about
\(2\times10^5\) degrees of freedom, and the \(20{,}000\) snapshots are split
chronologically, so that the test period follows the training period
(Appendix~\ref{app:details}).

Figure~\ref{fig:minimal3d}a shows the accuracy of each method from \(D=32\)
to \(2048\). OPINE improves on POD at every \(D\), by up to \(14\%\), and its
advantage grows with \(D\). The CNN autoencoder follows POD at small \(D\),
but from \(D=256\) its training and test errors both level off, whereas
OPINE's error keeps decreasing, with \(20\) to \(200\) times fewer
parameters than the CNN. In a wall-parallel plane
(Figure~\ref{fig:minimal3d}b), POD and the CNN smooth the near-wall streaks,
whereas OPINE preserves more of their pattern and of the smaller-scale
fluctuations.

Figure~\ref{fig:minimal3d}c,d shows the reconstructed vortices and streaks in three dimensions. Isosurfaces of the \(Q\)-criterion \citep{hunt1988eddies}, which mark
vortex cores, show that POD recovers only a few smoothed vortices, while the CNN and
OPINE recover larger sets. Isosurfaces of the streamwise fluctuation show the
low-speed streak that dominates the box: all methods recover it, but OPINE
reproduces more of its undulation and of the smaller high-speed regions near
the wall.

\subsection{Nested Reduced Coordinates}
\label{sec:nested}

In the comparisons above, OPINE was trained separately at each \(D\). Because
its POD basis is ordered, a single model trained at the largest \(D\) also
gives every smaller \(D\) by keeping only the leading modes
(Section~\ref{sec:training}; Q4). The prefixes are more accurate than POD at every \(D\) on Kolmogorov flow
and the minimal channel, and from \(D=3\) on the synthetic manifolds; on the minimal channel they come within \(4\%\) of separately trained models
from \(D=512\) onward (Table~\ref{tab:prefix_turb};
Appendix~\ref{app:synthetic_results}). As with POD, one training run thus gives accurate models of every size. For the minimal channel, the
prefixes of one model replace seven separately trained models, one for each
\(D\) from \(32\) to \(2048\). As in POD, the accuracy of each prefix in the
transformed coordinates is also known without any reconstruction: the mean
squared error of the prefix of dimension \(r\) on the training data equals
\(\operatorname{tr}C_z-\sum_{j\leq r}(\Lambda_Q)_{jj}\), with \(\Lambda_Q\) from Section~\ref{sec:training} in decreasing order, so the eigenvalues show how much variance each
smaller \(D\) retains.

\FloatBarrier
\section{Conclusion}

This paper makes three contributions. First, OPINE factors nonlinear
dimensionality reduction into an invertible change of coordinates and
an explicit, ordered POD projection, so that all information loss is confined to the projection and one model
gives reduced coordinates of every smaller dimension. Second, its training tracks the POD subspace with a
damped, aligned step of orthogonal iteration, which avoids differentiating
through an eigendecomposition, provably converges locally to the exact POD
subspace, and needs only minibatch products with an \(N\times K\) basis.
Third, from synthetic manifolds to three-dimensional turbulence, OPINE
improves on POD at every retained dimension above one, with far fewer
parameters than the CNN autoencoders; on the KS and Burgers dynamics it is more than two orders of magnitude more
accurate than autoencoders at the largest retained dimension, and the
ablations support the choice of the POD projection. OPINE is not designed to outperform unconstrained autoencoders on every
problem; its purpose is an interpretable reduction built only from an
invertible change of variables and POD, and it is most beneficial when the
states lie near a low-dimensional set that a linear subspace cannot capture. Its coordinates also remain
interpretable: decoded one at a time, they form phase-shifted pairs like POD
modes but sharper (Figure~\ref{fig:kolmogorov_latent_directions}). The same construction, an invertible
nonlinear map followed by an ordered POD projection, can be used wherever POD
is used, for example for compression, modal analysis, and reduced-order
modeling of other physical fields.

\paragraph{Limitations and future work.}
Training reduced dynamical models in OPINE
coordinates is a natural next step beyond the reconstruction studied here;
such models will need hyper-reduction because
\(f_\theta\) acts on the full field. The subspace could also be optimized
directly on the Grassmann manifold
\citep{edelman1998geometry,absil2008optimization} instead of by damped
orthogonal iteration. Projections constrained by symmetry, with POD
performed within Fourier blocks of a homogeneous direction, are a further
extension that we leave for future work. The analysis of the subspace update is local and does not cover
the joint training of the map and the basis (Appendix~\ref{app:pod}), and the flow datasets are simulations at moderate Reynolds numbers.

\clearpage
\section*{AI Use Statement}

Generative AI tools were used to assist with editing and polishing the
manuscript, including improving clarity, grammar, and presentation, and to help
draw parts of the schematics. The analysis of the subspace update in
Appendix~\ref{app:pod}, including the statements and proofs of
Theorem~\ref{thm:relaxed} and Proposition~\ref{prop:tracking}, was drafted
mainly with the help of generative AI; the authors checked every step of the
derivations. All AI-assisted content was reviewed and verified by the
authors, who take full responsibility for the final content of the paper.

\section*{Reproducibility Statement}

Section~\ref{sec:method} specifies the model and the training algorithm, and
Appendix~\ref{app:pod} states the assumptions of the analysis.
Appendices~\ref{app:synthetic_results} and~\ref{app:details} give the
architectures, optimizer settings, data generation, preprocessing, and data
splits of every experiment. Code and data-generation scripts will be released
upon publication.

\bibliography{preprint}

\begin{thebibliography}{53}
\providecommand{\natexlab}[1]{#1}
\providecommand{\url}[1]{\texttt{#1}}
\expandafter\ifx\csname urlstyle\endcsname\relax
  \providecommand{\doi}[1]{doi: #1}\else
  \providecommand{\doi}{doi: \begingroup \urlstyle{rm}\Url}\fi

\bibitem[Absil et~al.(2008)Absil, Mahony, and Sepulchre]{absil2008optimization}
P.-A. Absil, Robert Mahony, and Rodolphe Sepulchre.
\newblock \emph{Optimization Algorithms on Matrix Manifolds}.
\newblock Princeton University Press, 2008.

\bibitem[Behrmann et~al.(2019)Behrmann, Grathwohl, Chen, Duvenaud, and
  Jacobsen]{behrmann2019iresnet}
Jens Behrmann, Will Grathwohl, Ricky T.~Q. Chen, David Duvenaud, and
  J{\"o}rn-Henrik Jacobsen.
\newblock Invertible residual networks.
\newblock In \emph{International Conference on Machine Learning}, pp.\
  573--582, 2019.

\bibitem[Bekasov \& Murray(2020)Bekasov and Murray]{bekasov2020ordering}
Artur Bekasov and Iain Murray.
\newblock Ordering dimensions with nested dropout normalizing flows.
\newblock In \emph{ICML Workshop on Invertible Neural Networks, Normalizing
  Flows, and Explicit Likelihood Models}, 2020.

\bibitem[Berkooz et~al.(1993)Berkooz, Holmes, and Lumley]{berkooz1993proper}
Gal Berkooz, Philip Holmes, and John~L. Lumley.
\newblock The proper orthogonal decomposition in the analysis of turbulent
  flows.
\newblock \emph{Annual Review of Fluid Mechanics}, 25:\penalty0 539--575, 1993.
\newblock \doi{10.1146/annurev.fl.25.010193.002543}.

\bibitem[Brehmer \& Cranmer(2020)Brehmer and Cranmer]{brehmer2020flows}
Johann Brehmer and Kyle Cranmer.
\newblock Flows for simultaneous manifold learning and density estimation.
\newblock In \emph{Advances in Neural Information Processing Systems},
  volume~33, 2020.

\bibitem[Caterini et~al.(2021)Caterini, Loaiza-Ganem, Pleiss, and
  Cunningham]{caterini2021rectangular}
Anthony~L. Caterini, Gabriel Loaiza-Ganem, Geoff Pleiss, and John~P.
  Cunningham.
\newblock Rectangular flows for manifold learning.
\newblock In \emph{Advances in Neural Information Processing Systems},
  volume~34, 2021.

\bibitem[Champion et~al.(2019)Champion, Lusch, Kutz, and
  Brunton]{champion2019data}
Kathleen Champion, Bethany Lusch, J.~Nathan Kutz, and Steven~L. Brunton.
\newblock Data-driven discovery of coordinates and governing equations.
\newblock \emph{Proceedings of the National Academy of Sciences}, 116\penalty0
  (45):\penalty0 22445--22451, 2019.

\bibitem[Chen et~al.(2019)Chen, Behrmann, Duvenaud, and
  Jacobsen]{chen2019residual}
Ricky T.~Q. Chen, Jens Behrmann, David~K. Duvenaud, and J{\"o}rn-Henrik
  Jacobsen.
\newblock Residual flows for invertible generative modeling.
\newblock In \emph{Advances in Neural Information Processing Systems},
  volume~32, 2019.

\bibitem[Cunningham et~al.(2022)Cunningham, Cobb, and Jha]{cunningham2022pcf}
Edmond Cunningham, Adam~D. Cobb, and Susmit Jha.
\newblock Principal component flows.
\newblock In \emph{International Conference on Machine Learning}, 2022.

\bibitem[Davis \& Kahan(1970)Davis and Kahan]{davis1970rotation}
Chandler Davis and W.~M. Kahan.
\newblock The rotation of eigenvectors by a perturbation. {III}.
\newblock \emph{SIAM Journal on Numerical Analysis}, 7\penalty0 (1):\penalty0
  1--46, 1970.

\bibitem[Dinh et~al.(2017)Dinh, Sohl-Dickstein, and Bengio]{dinh2017realnvp}
Laurent Dinh, Jascha Sohl-Dickstein, and Samy Bengio.
\newblock Density estimation using {Real NVP}.
\newblock In \emph{International Conference on Learning Representations}, 2017.

\bibitem[Durkan et~al.(2019)Durkan, Bekasov, Murray, and
  Papamakarios]{durkan2019neural}
Conor Durkan, Artur Bekasov, Iain Murray, and George Papamakarios.
\newblock Neural spline flows.
\newblock In \emph{Advances in Neural Information Processing Systems},
  volume~32, 2019.

\bibitem[Edelman et~al.(1998)Edelman, Arias, and Smith]{edelman1998geometry}
Alan Edelman, Tom{\'a}s~A. Arias, and Steven~T. Smith.
\newblock The geometry of algorithms with orthogonality constraints.
\newblock \emph{SIAM Journal on Matrix Analysis and Applications}, 20\penalty0
  (2):\penalty0 303--353, 1998.

\bibitem[Eivazi et~al.(2022)Eivazi, Le~Clainche, Hoyas, and
  Vinuesa]{eivazi2022towards}
Hamidreza Eivazi, Soledad Le~Clainche, Sergio Hoyas, and Ricardo Vinuesa.
\newblock Towards extraction of orthogonal and parsimonious non-linear modes
  from turbulent flows.
\newblock \emph{Expert Systems with Applications}, 202:\penalty0 117038, 2022.

\bibitem[Fan(1949)]{fan1949theorem}
Ky~Fan.
\newblock On a theorem of {W}eyl concerning eigenvalues of linear
  transformations {I}.
\newblock \emph{Proceedings of the National Academy of Sciences}, 35\penalty0
  (11):\penalty0 652--655, 1949.

\bibitem[Fresca \& Manzoni(2022)Fresca and Manzoni]{fresca2022poddrom}
Stefania Fresca and Andrea Manzoni.
\newblock {POD-DL-ROM}: Enhancing deep learning-based reduced order models for
  nonlinear parametrized {PDEs} by proper orthogonal decomposition.
\newblock \emph{Computer Methods in Applied Mechanics and Engineering},
  388:\penalty0 114181, 2022.
\newblock \doi{10.1016/j.cma.2021.114181}.

\bibitem[Fukami et~al.(2020)Fukami, Nakamura, and
  Fukagata]{fukami2020hierarchical}
Kai Fukami, Taichi Nakamura, and Koji Fukagata.
\newblock Convolutional neural network based hierarchical autoencoder for
  nonlinear mode decomposition of fluid field data.
\newblock \emph{Physics of Fluids}, 32\penalty0 (9):\penalty0 095110, 2020.
\newblock \doi{10.1063/5.0020721}.

\bibitem[Geelen et~al.(2023)Geelen, Wright, and Willcox]{geelen2023quadratic}
Rudy Geelen, Stephen Wright, and Karen Willcox.
\newblock Operator inference for non-intrusive model reduction with quadratic
  manifolds.
\newblock \emph{Computer Methods in Applied Mechanics and Engineering},
  403:\penalty0 115717, 2023.

\bibitem[Golub \& Van~Loan(2013)Golub and Van~Loan]{golub2013matrix}
Gene~H. Golub and Charles~F. Van~Loan.
\newblock \emph{Matrix Computations}.
\newblock Johns Hopkins University Press, Baltimore, MD, 4th edition, 2013.

\bibitem[Greif \& Urban(2019)Greif and Urban]{greif2019decay}
Constantin Greif and Karsten Urban.
\newblock Decay of the {K}olmogorov $n$-width for wave problems.
\newblock \emph{Applied Mathematics Letters}, 96:\penalty0 216--222, 2019.
\newblock \doi{10.1016/j.aml.2019.05.013}.

\bibitem[Hinton \& Salakhutdinov(2006)Hinton and
  Salakhutdinov]{hinton2006reducing}
Geoffrey~E. Hinton and Ruslan~R. Salakhutdinov.
\newblock Reducing the dimensionality of data with neural networks.
\newblock \emph{Science}, 313\penalty0 (5786):\penalty0 504--507, 2006.

\bibitem[Holmes et~al.(2012)Holmes, Lumley, Berkooz, and
  Rowley]{holmes2012turbulence}
Philip Holmes, John~L. Lumley, Gal Berkooz, and Clarence~W. Rowley.
\newblock \emph{Turbulence, Coherent Structures, Dynamical Systems and
  Symmetry}.
\newblock Cambridge University Press, 2nd edition, 2012.

\bibitem[Huang et~al.(2018)Huang, Krueger, Lacoste, and
  Courville]{huang2018neural}
Chin-Wei Huang, David Krueger, Alexandre Lacoste, and Aaron Courville.
\newblock Neural autoregressive flows.
\newblock In \emph{International Conference on Machine Learning}, pp.\
  2078--2087, 2018.

\bibitem[Hunt et~al.(1988)Hunt, Wray, and Moin]{hunt1988eddies}
Julian C.~R. Hunt, Alan~A. Wray, and Parviz Moin.
\newblock Eddies, streams, and convergence zones in turbulent flows.
\newblock In \emph{Proceedings of the Summer Program, Center for Turbulence
  Research}, pp.\  193--208, 1988.
\newblock CTR-S88.

\bibitem[Ionescu et~al.(2015)Ionescu, Vantzos, and
  Sminchisescu]{ionescu2015matrix}
Catalin Ionescu, Orestis Vantzos, and Cristian Sminchisescu.
\newblock Matrix backpropagation for deep networks with structured layers.
\newblock In \emph{IEEE International Conference on Computer Vision}, 2015.

\bibitem[Jim{\'e}nez \& Moin(1991)Jim{\'e}nez and Moin]{jimenez1991minimal}
Javier Jim{\'e}nez and Parviz Moin.
\newblock The minimal flow unit in near-wall turbulence.
\newblock \emph{Journal of Fluid Mechanics}, 225:\penalty0 213--240, 1991.
\newblock \doi{10.1017/S0022112091002033}.

\bibitem[Kassam \& Trefethen(2005)Kassam and Trefethen]{kassam2005fourth}
Aly-Khan Kassam and Lloyd~N. Trefethen.
\newblock Fourth-order time-stepping for stiff {PDE}s.
\newblock \emph{SIAM Journal on Scientific Computing}, 26\penalty0
  (4):\penalty0 1214--1233, 2005.

\bibitem[Kim et~al.(1987)Kim, Moin, and Moser]{kim1987turbulence}
John Kim, Parviz Moin, and Robert Moser.
\newblock Turbulence statistics in fully developed channel flow at low reynolds
  number.
\newblock \emph{Journal of Fluid Mechanics}, 177:\penalty0 133--166, 1987.
\newblock \doi{10.1017/S0022112087000892}.

\bibitem[Kingma \& Dhariwal(2018)Kingma and Dhariwal]{kingma2018glow}
Diederik~P. Kingma and Prafulla Dhariwal.
\newblock Glow: Generative flow with invertible 1x1 convolutions.
\newblock In \emph{Advances in Neural Information Processing Systems},
  volume~31, 2018.

\bibitem[Kingma \& Welling(2014)Kingma and Welling]{kingma2014vae}
Diederik~P. Kingma and Max Welling.
\newblock Auto-encoding variational {B}ayes.
\newblock In \emph{International Conference on Learning Representations}, 2014.

\bibitem[Kochkov et~al.(2021)Kochkov, Smith, Alieva, Wang, Brenner, and
  Hoyer]{kochkov2021machine}
Dmitrii Kochkov, Jamie~A. Smith, Ayya Alieva, Qing Wang, Michael~P. Brenner,
  and Stephan Hoyer.
\newblock Machine learning-accelerated computational fluid dynamics.
\newblock \emph{Proceedings of the National Academy of Sciences}, 118\penalty0
  (21):\penalty0 e2101784118, 2021.
\newblock \doi{10.1073/pnas.2101784118}.

\bibitem[Kothari et~al.(2021)Kothari, Khorashadizadeh, de~Hoop, and
  Dokmani{\'c}]{kothari2021trumpets}
Konik Kothari, AmirEhsan Khorashadizadeh, Maarten de~Hoop, and Ivan
  Dokmani{\'c}.
\newblock Trumpets: Injective flows for inference and inverse problems.
\newblock In \emph{Conference on Uncertainty in Artificial Intelligence}, 2021.

\bibitem[Kramer(1991)]{kramer1991nonlinear}
Mark~A. Kramer.
\newblock Nonlinear principal component analysis using autoassociative neural
  networks.
\newblock \emph{AIChE Journal}, 37\penalty0 (2):\penalty0 233--243, 1991.
\newblock \doi{10.1002/aic.690370209}.

\bibitem[Kusupati et~al.(2022)Kusupati, Bhatt, Rege, Wallingford, Sinha,
  Ramanujan, Howard-Snyder, Chen, Kakade, Jain, and
  Farhadi]{kusupati2022matryoshka}
Aditya Kusupati, Gantavya Bhatt, Aniket Rege, Matthew Wallingford, Aditya
  Sinha, Vivek Ramanujan, William Howard-Snyder, Kaifeng Chen, Sham Kakade,
  Prateek Jain, and Ali Farhadi.
\newblock Matryoshka representation learning.
\newblock In \emph{Advances in Neural Information Processing Systems},
  volume~35, 2022.

\bibitem[Lee \& Carlberg(2020)Lee and Carlberg]{lee2020model}
Kookjin Lee and Kevin~T. Carlberg.
\newblock Model reduction of dynamical systems on nonlinear manifolds using
  deep convolutional autoencoders.
\newblock \emph{Journal of Computational Physics}, 404:\penalty0 108973, 2020.
\newblock \doi{10.1016/j.jcp.2019.108973}.

\bibitem[Li \& Hooi(2022)Li and Hooi]{li2022neural}
Shen Li and Bryan Hooi.
\newblock Neural {PCA} for flow-based representation learning.
\newblock In \emph{International Joint Conference on Artificial Intelligence},
  pp.\  3229--3235, 2022.
\newblock \doi{10.24963/ijcai.2022/448}.

\bibitem[Linot \& Graham(2020)Linot and Graham]{linot2020deep}
Alec~J. Linot and Michael~D. Graham.
\newblock Deep learning to discover and predict dynamics on an inertial
  manifold.
\newblock \emph{Physical Review E}, 101:\penalty0 062209, 2020.

\bibitem[Loshchilov \& Hutter(2019)Loshchilov and Hutter]{loshchilov2019adamw}
Ilya Loshchilov and Frank Hutter.
\newblock Decoupled weight decay regularization.
\newblock In \emph{International Conference on Learning Representations}, 2019.

\bibitem[Lumley(1970)]{lumley1970stochastic}
John~L. Lumley.
\newblock \emph{Stochastic Tools in Turbulence}.
\newblock Academic Press, 1970.

\bibitem[Lusch et~al.(2018)Lusch, Kutz, and Brunton]{lusch2018deep}
Bethany Lusch, J.~Nathan Kutz, and Steven~L. Brunton.
\newblock Deep learning for universal linear embeddings of nonlinear dynamics.
\newblock \emph{Nature Communications}, 9:\penalty0 4950, 2018.

\bibitem[Maulik et~al.(2021)Maulik, Lusch, and Balaprakash]{maulik2021reduced}
Romit Maulik, Bethany Lusch, and Prasanna Balaprakash.
\newblock Reduced-order modeling of advection-dominated systems with recurrent
  neural networks and convolutional autoencoders.
\newblock \emph{Physics of Fluids}, 33\penalty0 (3):\penalty0 037106, 2021.

\bibitem[Mika et~al.(1999)Mika, Sch{\"o}lkopf, Smola, M{\"u}ller, Scholz, and
  R{\"a}tsch]{mika1999kernel}
Sebastian Mika, Bernhard Sch{\"o}lkopf, Alex Smola, Klaus-Robert M{\"u}ller,
  Matthias Scholz, and Gunnar R{\"a}tsch.
\newblock Kernel {PCA} and de-noising in feature spaces.
\newblock In \emph{Advances in Neural Information Processing Systems},
  volume~11, 1999.

\bibitem[Milano \& Koumoutsakos(2002)Milano and Koumoutsakos]{milano2002neural}
Michele Milano and Petros Koumoutsakos.
\newblock Neural network modeling for near wall turbulent flow.
\newblock \emph{Journal of Computational Physics}, 182\penalty0 (1):\penalty0
  1--26, 2002.

\bibitem[M{\"u}ller et~al.(2019)M{\"u}ller, McWilliams, Rousselle, Gross, and
  Nov{\'a}k]{muller2019neural}
Thomas M{\"u}ller, Brian McWilliams, Fabrice Rousselle, Markus Gross, and Jan
  Nov{\'a}k.
\newblock Neural importance sampling.
\newblock \emph{ACM Transactions on Graphics}, 38\penalty0 (5):\penalty0
  145:1--145:19, 2019.

\bibitem[Murata et~al.(2020)Murata, Fukami, and Fukagata]{murata2020nonlinear}
Takaaki Murata, Kai Fukami, and Koji Fukagata.
\newblock Nonlinear mode decomposition with convolutional neural networks for
  fluid dynamics.
\newblock \emph{Journal of Fluid Mechanics}, 882:\penalty0 A13, 2020.
\newblock \doi{10.1017/jfm.2019.822}.

\bibitem[Oja(1982)]{oja1982simplified}
Erkki Oja.
\newblock A simplified neuron model as a principal component analyzer.
\newblock \emph{Journal of Mathematical Biology}, 15\penalty0 (3):\penalty0
  267--273, 1982.
\newblock \doi{10.1007/BF00275687}.

\bibitem[Reiss et~al.(2018)Reiss, Schulze, Sesterhenn, and
  Mehrmann]{reiss2018shifted}
Julius Reiss, Philipp Schulze, J{\"o}rn Sesterhenn, and Volker Mehrmann.
\newblock The shifted proper orthogonal decomposition: A mode decomposition for
  multiple transport phenomena.
\newblock \emph{SIAM Journal on Scientific Computing}, 40\penalty0
  (3):\penalty0 A1322--A1344, 2018.

\bibitem[Rippel et~al.(2014)Rippel, Gelbart, and Adams]{rippel2014nested}
Oren Rippel, Michael~A. Gelbart, and Ryan~P. Adams.
\newblock Learning ordered representations with nested dropout.
\newblock In \emph{International Conference on Machine Learning}, pp.\
  1746--1754, 2014.

\bibitem[Sch{\"o}lkopf et~al.(1998)Sch{\"o}lkopf, Smola, and
  M{\"u}ller]{scholkopf1998kpca}
Bernhard Sch{\"o}lkopf, Alexander Smola, and Klaus-Robert M{\"u}ller.
\newblock Nonlinear component analysis as a kernel eigenvalue problem.
\newblock \emph{Neural Computation}, 10\penalty0 (5):\penalty0 1299--1319,
  1998.

\bibitem[Taddei(2020)]{taddei2020registration}
Tommaso Taddei.
\newblock A registration method for model order reduction: Data compression and
  geometry reduction.
\newblock \emph{SIAM Journal on Scientific Computing}, 42\penalty0
  (2):\penalty0 A997--A1027, 2020.

\bibitem[Taira et~al.(2017)Taira, Brunton, Dawson, Rowley, Colonius, McKeon,
  Schmidt, Gordeyev, Theofilis, and Ukeiley]{taira2017modal}
Kunihiko Taira, Steven~L. Brunton, Scott T.~M. Dawson, Clarence~W. Rowley, Tim
  Colonius, Beverley~J. McKeon, Oliver~T. Schmidt, Stanislav Gordeyev,
  Vassilios Theofilis, and Lawrence~S. Ukeiley.
\newblock Modal analysis of fluid flows: An overview.
\newblock \emph{AIAA Journal}, 55\penalty0 (12):\penalty0 4013--4041, 2017.

\bibitem[Wang et~al.(2019)Wang, Dang, Hu, Fua, and
  Salzmann]{wang2019backpropagation}
Wei Wang, Zheng Dang, Yinlin Hu, Pascal Fua, and Mathieu Salzmann.
\newblock Backpropagation-friendly eigendecomposition.
\newblock In \emph{Advances in Neural Information Processing Systems},
  volume~32, 2019.

\bibitem[Yang(1995)]{yang1995projection}
Bin Yang.
\newblock Projection approximation subspace tracking.
\newblock \emph{IEEE Transactions on Signal Processing}, 43\penalty0
  (1):\penalty0 95--107, 1995.
\newblock \doi{10.1109/78.365290}.

\end{thebibliography}
\bibliographystyle{preprint}

\clearpage
\appendix

\section{Analysis of the Subspace Update}
\label{app:pod}

During training, OPINE does not recompute the POD basis exactly. After each
round of gradient updates of \(f_\theta\), it moves the basis only a fraction
\(\alpha\) of the way toward one step of orthogonal iteration,
Eq.~\plaineqref{eq:q_update}. This appendix shows that this damped update has
the right target. For a fixed map with a spectral gap, it converges locally to the exact POD
subspace, the only nearby fixed
point, for every \(\alpha\in(0,1]\) (Theorem~\ref{thm:relaxed}). While
\(f_\theta\) changes, it trails that subspace by a bounded angle
(Proposition~\ref{prop:tracking}). The method itself does not require the
exact basis: every quantity in Section~\ref{sec:method} is defined for any
orthonormal \(Q_K\), which only makes the transformed projection error larger
than its minimum. We use two textbook facts without proof: for fixed
\(\theta\), the minimizers of Eq.~\plaineqref{eq:pod} span a leading
\(K\)-dimensional eigenspace of \(C_z\) \citep{fan1949theorem,holmes2012turbulence},
and plain orthogonal iteration, the case \(\alpha=1\), converges to it at rate
\(\lambda_{K+1}/\lambda_K\) \citep[Sec.~8.2.4]{golub2013matrix}. The joint, nonconvex optimization of \(f_\theta\) and the basis is beyond the
scope of this analysis.

\paragraph{Notation.}
We write \(Q\) for the basis \(Q_K\) and \(\widetilde Q\) for
\(\widetilde Q_K\) of Section~\ref{sec:training}, and \(\|\cdot\|\) for the
spectral norm of a matrix. Let \(C_z=V\Lambda V^\top\) with \(V=[V_K,V_\perp]\) orthogonal,
\(\Lambda=\operatorname{diag}(\Lambda_K,\Lambda_\perp)\), eigenvalues
\(\lambda_1\geq\cdots\geq\lambda_N\geq0\), and \(\lambda_K>\lambda_{K+1}\). For
\(Q\in\mathbb{R}^{N\times K}\) with orthonormal columns and \(V_K^\top Q\)
nonsingular, write \(\operatorname{span}(Q)=\operatorname{span}(V_K+V_\perp T)\)
with \(T=T(Q)=V_\perp^\top Q(V_K^\top Q)^{-1}\); then \(\|\tan\Theta\|=\|T\|\) for
the principal angles \(\Theta\) between \(\operatorname{span}(Q)\) and
\(\operatorname{span}(V_K)\), and \(Q_T=X_T(I+T^\top T)^{-1/2}\) with
\(X_T=V_K+V_\perp T\) is an orthonormal basis of that span
\citep[Sec.~6.4.3]{golub2013matrix}. The update of Eq.~\plaineqref{eq:q_update}
is \(\Phi_\alpha(Q)=\operatorname{qf}[(1-\alpha)Q+\alpha\widetilde QR]\), with
\(\widetilde Q=\operatorname{qf}(C_zQ)\) and \(R\) the Procrustes rotation of
Eq.~\plaineqref{eq:power}.

\begin{theorem}[Convergence of the aligned, damped update]
\label{thm:relaxed}
Let \(\alpha\in(0,1]\) and
\(\rho(\alpha)=1-\alpha\,(1-\lambda_{K+1}/\lambda_K)<1\). Near
\(\operatorname{span}(V_K)\), the subspace \(\operatorname{span}(\Phi_\alpha(Q))\)
depends only on \(\operatorname{span}(Q)\), and there are constants
\(\varepsilon,c>0\) such that \(\|T(Q)\|\leq\varepsilon\) implies
\[
\bigl\|T(\Phi_\alpha(Q))-\mathcal A_\alpha(T(Q))\bigr\|\leq c\,\|T(Q)\|^2,
\qquad
\mathcal A_\alpha(T)=(1-\alpha)T+\alpha\Lambda_\perp T\Lambda_K^{-1},
\qquad
\|\mathcal A_\alpha(T)\|\leq\rho(\alpha)\|T\|.
\]
Consequently, for every \(\bar\rho\in(\rho(\alpha),1)\), iterates started close
enough to \(\operatorname{span}(V_K)\) satisfy
\(\|\tan\Theta_t\|\leq\bar\rho^{\,t}\|\tan\Theta_0\|\), where \(\Theta_t\) are the
principal angles after \(t\) updates, and
\(\operatorname{span}(V_K)\), the exact POD subspace, is the only fixed point
of \(\Phi_\alpha\) in this neighborhood, for every \(\alpha\).
\end{theorem}

\begin{proof}
\emph{Span only.} Replacing \(Q\) by \(Q\Omega\) with \(\Omega\) orthogonal leaves
\(\operatorname{span}(\widetilde Q)\) unchanged, and \(\widetilde QR\) is the
orthonormal basis \(Q'\) of that span closest to \(Q\) in Frobenius norm, which
is unique near the fixed point, where \(\widetilde Q^\top Q\) is nonsingular.
Since \(\|Q'\Omega-Q\Omega\|_F=\|Q'-Q\|_F\), the basis closest to \(Q\Omega\) is
\(Q'\Omega\), and the argument of \(\operatorname{qf}\) becomes
\([(1-\alpha)Q+\alpha Q']\Omega\), with the
same span. We may therefore take \(Q=Q_T\).

\emph{Linearization.} Since
\(C_zX_T=V_K\Lambda_K+V_\perp\Lambda_\perp T=X_{\widetilde T}\Lambda_K\) with
\(\widetilde T=\Lambda_\perp T\Lambda_K^{-1}\), the span of \(\widetilde Q\) is
that of \(X_{\widetilde T}\), so \(Q'=Q_{\widetilde T}H\), with \(H\) the orthogonal
polar factor of
\(Q_{\widetilde T}^\top Q_T=(I+\widetilde T^\top\widetilde T)^{-1/2}(I+\widetilde T^\top T)(I+T^\top T)^{-1/2}=I+\mathcal O(\|T\|^2)\).
Hence \(H=I+\mathcal O(\|T\|^2)\), \(Q'=V_K+V_\perp\widetilde T+\mathcal O(\|T\|^2)\), and
\(Q_T=V_K+V_\perp T+\mathcal O(\|T\|^2)\), so
\(Y=(1-\alpha)Q_T+\alpha Q'=V_K(I+\mathcal O(\|T\|^2))+V_\perp(\mathcal A_\alpha(T)+\mathcal O(\|T\|^2))\)
and \(T(\Phi_\alpha(Q))=V_\perp^\top Y(V_K^\top Y)^{-1}=\mathcal A_\alpha(T)+\mathcal O(\|T\|^2)\).
All maps involved are smooth near \(T=0\), which gives \(\varepsilon\) and \(c\).

\emph{Contraction and fixed point.}
\(\|\mathcal A_\alpha(T)\|\leq(1-\alpha)\|T\|+\alpha(\lambda_{K+1}/\lambda_K)\|T\|=\rho(\alpha)\|T\|\).
If \(\|T(Q)\|\leq\varepsilon'\leq\varepsilon\) with
\(\rho(\alpha)+c\varepsilon'\leq\bar\rho\), then
\(\|T(\Phi_\alpha(Q))\|\leq\bar\rho\|T(Q)\|\), and induction gives the rate. A
fixed span in this neighborhood satisfies \(\|T\|\leq\bar\rho\|T\|\), hence
\(T=0\); conversely, \(\Phi_\alpha(V_K)=V_K\) because \(C_zV_K=V_K\Lambda_K\).
\end{proof}

The alignment is necessary. Without it, \(\widetilde Q\) may differ from \(Q\)
by an arbitrary rotation even at the fixed point: if the columns of \(Q\) are
eigenvectors of \(C_z\), a QR factorization without sign normalization may
return \(\widetilde Q=-Q\), and then \((1-\alpha)Q+\alpha\widetilde Q=(1-2\alpha)Q\)
is rank deficient at \(\alpha=1/2\).

\begin{proposition}[Tracking a moving covariance]
\label{prop:tracking}
Let \(C_t\) be the transformed covariance at round \(t\), with gap
\(\lambda_K(C_t)-\lambda_{K+1}(C_t)\geq g>0\) and drift
\(\|C_{t+1}-C_t\|\leq\delta\leq g/2\), and let \(\vartheta_t\) be the largest
principal angle between \(\operatorname{span}(Q_K)\) after round \(t\) and the leading
eigenspace of \(C_t\). If each update contracts this angle by a factor
\(\bar\rho<1\) whenever it is at most \(\varepsilon\), as
Theorem~\ref{thm:relaxed} provides for \(\bar\rho>\rho(\alpha)\), and if
\(\vartheta_0\leq\varepsilon\) and \(\beta=\arcsin(2\delta/g)\leq(1-\bar\rho)\varepsilon\),
then \(\vartheta_t\leq\varepsilon\) for all \(t\) and
\(\limsup_t\vartheta_t\leq\beta/(1-\bar\rho)\approx2\delta/\bigl(g\,\alpha\,(1-\lambda_{K+1}/\lambda_K)\bigr)\).
\end{proposition}

\begin{proof}
By Weyl's inequality and the Davis--Kahan \(\sin\Theta\) theorem
\citep{davis1970rotation}, consecutive leading eigenspaces differ by an angle
of at most \(\arcsin(\delta/(g-\delta))\leq\beta\). The largest principal angle
is a metric on subspaces of equal dimension, so
\(\vartheta_{t+1}\leq\bar\rho\,\vartheta_t+\beta\), which stays below
\(\varepsilon\) by induction and unrolls to the bound.
\end{proof}

A larger \(\alpha\) therefore tracks a moving covariance more closely, whereas a
smaller \(\alpha\) changes the bottleneck, and hence the objective seen by the
gradient updates, more gradually; Appendix~\ref{app:synthetic_results} compares
several choices.

\section{One-Dimensional Study}
\label{app:synthetic_results}

The ablations of Section~\ref{sec:synthetic} and the experiments of
Section~\ref{sec:onedim} concern scalar fields on a periodic one-dimensional
grid and use a compact training configuration, described below, that differs
from that of the other experiments (Appendix~\ref{app:details}).

\paragraph{Implementation.}
The spatial coupling follows Section~\ref{sec:coupling}: a space-to-depth
rearrangement with factor \(4\), followed by \(L\) blocks of invertible
channel mixing and affine coupling, whose conditioners are three-layer
circularly padded convolutional networks of width \(w\) with kernel size \(5\)
and log-scales bounded by \(2\tanh(\cdot/2)\). The spectral coupling is an
exactly translation-equivariant alternative to the space-to-depth
rearrangement, which breaks translation symmetry. Each spectral-split coupling
layer partitions the wavenumbers into sets \(\mathcal K_a\) and \(\mathcal K_b\), keeps the
band-limited part \(a\) of the field, and updates the Fourier coefficients of
the other part as \(\hat b'=\hat b\odot\exp(\sigma_b(|\hat a|))+m_b\odot\mathcal F\,\tau_b(a)\),
where \(\sigma_b\) returns one real log-scale per wavenumber in \(\mathcal K_b\) from the Fourier
moduli of \(a\), \(\tau_b\) is a circularly padded convolutional network, and
\(m_b\) masks the wavenumbers in \(\mathcal K_b\). Its inverse is available in closed form,
and because the scales depend only on Fourier moduli and \(\tau_b\) is a circular
convolution, the layer commutes with grid shifts. Successive layers alternate
between partitions by \(k\bmod4\) and by \(k<4\), which divide both odd and even
harmonics between the two halves, whereas a partition into even and odd wavenumbers
leaves one half empty for data with only odd harmonics. Unless stated otherwise, \(L=4\) and \(w=32\).
The bottleneck is initialized with the exact POD of the transformed training
data and updated after each of \(20\) rounds of \(150\) AdamW steps (learning
rate \(10^{-3}\), cosine schedule, gradient clipping at \(1\)) with
\(\alpha=0.03\), and the checkpoint with the lowest validation NMSE is
reported. The CNN autoencoder has four stride-\(2\) circularly padded
convolutions and a mirrored decoder, and the MLP autoencoder has two hidden layers
of width \(256\) in the encoder and in the decoder. Both are trained for the
same \(3000\) steps. This study uses three seeds.

\paragraph{Datasets.}
The four synthetic families, defined in Section~\ref{sec:synthetic}, have
\(512\) training, \(256\) validation, and \(2000\) test samples each, with all
parameters drawn uniformly. The states of Pulse form a circle, those of Torus a
torus, and those of Pulse width and Front a cylinder. KS data are generated with ETDRK4 \citep{kassam2005fourth} on \(64\)
points with time step \(0.25\), and Burgers data with ETDRK4 on \(256\) points,
viscosity \(0.07\), forcing of amplitude \(0.5\) that is white in time and
acts on wavenumbers \(1\) to \(3\), and
storage on \(128\) points. Each system provides eight independent trajectories
of \(1000\) snapshots (spacing \(1\) for KS and \(0.5\) for Burgers, after a
transient), split by trajectory into six for training, one for validation, and
one for testing, and normalized by the training standard deviation.

\paragraph{Full results.}
Table~\ref{tab:onedim_full} reports all retained dimensions. Three patterns
stand out. First, OPINE improves on POD at every \(D\) above one, with error reductions of more than three orders of magnitude for the spatial coupling and four for the
spectral coupling. At \(D=1\), no method can be exact, because each
synthetic dataset contains a closed loop, which cannot be mapped continuously
and injectively onto a line. OPINE, whose encoder is an invertible map
followed by a linear projection, improves on POD only modestly there, whereas
the autoencoders reach errors of two to seven percent. Second,
for Torus, the errors of OPINE and the autoencoders all drop sharply at \(D=4\),
although a torus can be embedded in three dimensions. Third, the
CNN autoencoder is more accurate on Pulse and Pulse width, which contain a single traveling pulse, whereas OPINE is more accurate than both autoencoders on Front from
\(D=4\) and at least as accurate as them on KS and Burgers from \(D=8\)
onward. The spectral coupling is exactly translation-equivariant but less accurate
than the spatial coupling on KS and Burgers.

\begin{table}[!ht]
 \centering
 \footnotesize
 \setlength{\tabcolsep}{4pt}
 \caption{Median test NMSE of the one-dimensional study.
 OPINE uses the spatial coupling, OPINE-spectral the spectral-split
 coupling described above. Best in bold; ``--'' marks runs
 not completed.}
 \label{tab:onedim_full}
 \vspace{3pt}
 \begin{tabular}{lrccccc}
 \toprule
 Dataset & $D$ & POD & CNN & MLP & OPINE & OPINE-spectral \\
 \midrule
Torus & $1$ & $3.8{\times}10^{-1}$ & $4.9{\times}10^{-2}$ & $\mathbf{4.1{\times}10^{-2}}$ & $3.8{\times}10^{-1}$ & $3.7{\times}10^{-1}$ \\
 & $2$ & $5.2{\times}10^{-2}$ & $\mathbf{1.2{\times}10^{-2}}$ & $1.4{\times}10^{-2}$ & $1.3{\times}10^{-2}$ & $1.7{\times}10^{-2}$ \\
 & $3$ & $2.8{\times}10^{-2}$ & $1.1{\times}10^{-3}$ & $\mathbf{7.6{\times}10^{-4}}$ & $6.0{\times}10^{-3}$ & $6.3{\times}10^{-3}$ \\
 & $4$ & $4.8{\times}10^{-3}$ & $\mathbf{1.6{\times}10^{-5}}$ & $2.3{\times}10^{-5}$ & $1.9{\times}10^{-5}$ & $2.6{\times}10^{-5}$ \\
 & $6$ & $1.2{\times}10^{-3}$ & $1.8{\times}10^{-5}$ & $2.4{\times}10^{-5}$ & $1.5{\times}10^{-5}$ & $\mathbf{5.5{\times}10^{-6}}$ \\
\midrule
Pulse & $1$ & $7.3{\times}10^{-1}$ & $\mathbf{3.1{\times}10^{-2}}$ & $5.8{\times}10^{-2}$ & $4.9{\times}10^{-1}$ & $7.0{\times}10^{-1}$ \\
 & $2$ & $6.0{\times}10^{-1}$ & $\mathbf{5.0{\times}10^{-6}}$ & $4.9{\times}10^{-5}$ & $6.1{\times}10^{-4}$ & $4.3{\times}10^{-4}$ \\
 & $3$ & $4.9{\times}10^{-1}$ & $\mathbf{4.4{\times}10^{-6}}$ & $3.2{\times}10^{-5}$ & $6.7{\times}10^{-4}$ & $3.9{\times}10^{-4}$ \\
 & $4$ & $3.8{\times}10^{-1}$ & $\mathbf{5.0{\times}10^{-6}}$ & $3.1{\times}10^{-5}$ & $1.3{\times}10^{-4}$ & $2.1{\times}10^{-4}$ \\
 & $6$ & $2.3{\times}10^{-1}$ & $\mathbf{3.9{\times}10^{-6}}$ & $2.8{\times}10^{-5}$ & $4.7{\times}10^{-5}$ & $4.8{\times}10^{-5}$ \\
\midrule
Pulse width & $1$ & $6.4{\times}10^{-1}$ & $7.1{\times}10^{-2}$ & $\mathbf{6.6{\times}10^{-2}}$ & $5.2{\times}10^{-1}$ & $5.8{\times}10^{-1}$ \\
 & $2$ & $4.7{\times}10^{-1}$ & $\mathbf{4.8{\times}10^{-5}}$ & $1.8{\times}10^{-3}$ & $1.1{\times}10^{-2}$ & $1.0{\times}10^{-3}$ \\
 & $3$ & $3.6{\times}10^{-1}$ & $\mathbf{5.1{\times}10^{-5}}$ & $7.4{\times}10^{-4}$ & $1.9{\times}10^{-2}$ & $9.7{\times}10^{-4}$ \\
 & $4$ & $2.5{\times}10^{-1}$ & $\mathbf{4.4{\times}10^{-5}}$ & $5.6{\times}10^{-4}$ & $6.8{\times}10^{-4}$ & $1.5{\times}10^{-4}$ \\
 & $6$ & $1.3{\times}10^{-1}$ & $\mathbf{3.3{\times}10^{-5}}$ & $4.9{\times}10^{-4}$ & $2.2{\times}10^{-4}$ & $4.0{\times}10^{-5}$ \\
\midrule
Front & $1$ & $5.3{\times}10^{-1}$ & $\mathbf{2.4{\times}10^{-2}}$ & $3.4{\times}10^{-2}$ & $5.1{\times}10^{-1}$ & $5.1{\times}10^{-1}$ \\
 & $2$ & $6.2{\times}10^{-2}$ & $2.1{\times}10^{-5}$ & $1.6{\times}10^{-4}$ & $2.5{\times}10^{-5}$ & $\mathbf{1.1{\times}10^{-5}}$ \\
 & $3$ & $3.6{\times}10^{-2}$ & $2.1{\times}10^{-5}$ & $9.0{\times}10^{-5}$ & $2.3{\times}10^{-5}$ & $\mathbf{1.1{\times}10^{-5}}$ \\
 & $4$ & $1.0{\times}10^{-2}$ & $2.5{\times}10^{-5}$ & $7.8{\times}10^{-5}$ & $6.0{\times}10^{-6}$ & $\mathbf{6.5{\times}10^{-7}}$ \\
 & $6$ & $2.1{\times}10^{-3}$ & $1.5{\times}10^{-5}$ & $6.9{\times}10^{-5}$ & $2.3{\times}10^{-6}$ & $\mathbf{1.2{\times}10^{-7}}$ \\
\midrule
KS & $2$ & $5.7{\times}10^{-1}$ & $2.7{\times}10^{-1}$ & $\mathbf{2.4{\times}10^{-1}}$ & $4.2{\times}10^{-1}$ & $4.3{\times}10^{-1}$ \\
 & $4$ & $2.6{\times}10^{-1}$ & $4.9{\times}10^{-2}$ & $5.7{\times}10^{-2}$ & $\mathbf{2.3{\times}10^{-2}}$ & $4.1{\times}10^{-2}$ \\
 & $6$ & $9.1{\times}10^{-2}$ & $\mathbf{1.2{\times}10^{-3}}$ & $8.3{\times}10^{-3}$ & $2.5{\times}10^{-3}$ & $2.3{\times}10^{-2}$ \\
 & $8$ & $3.1{\times}10^{-2}$ & $1.1{\times}10^{-3}$ & $1.6{\times}10^{-4}$ & $\mathbf{3.8{\times}10^{-5}}$ & $8.9{\times}10^{-4}$ \\
 & $12$ & $1.4{\times}10^{-3}$ & $1.0{\times}10^{-3}$ & $1.5{\times}10^{-4}$ & $\mathbf{3.2{\times}10^{-5}}$ & $6.4{\times}10^{-5}$ \\
 & $16$ & $1.2{\times}10^{-4}$ & $9.7{\times}10^{-4}$ & $1.5{\times}10^{-4}$ & $\mathbf{1.7{\times}10^{-7}}$ & $2.8{\times}10^{-6}$ \\
\midrule
Burgers & $4$ & $2.3{\times}10^{-1}$ & $\mathbf{1.5{\times}10^{-1}}$ & $1.6{\times}10^{-1}$ & $1.5{\times}10^{-1}$ & $1.8{\times}10^{-1}$ \\
 & $8$ & $5.4{\times}10^{-2}$ & $9.1{\times}10^{-3}$ & $1.3{\times}10^{-2}$ & $\mathbf{9.1{\times}10^{-3}}$ & $2.1{\times}10^{-2}$ \\
 & $16$ & $1.2{\times}10^{-2}$ & $8.0{\times}10^{-4}$ & $2.9{\times}10^{-3}$ & $\mathbf{6.8{\times}10^{-5}}$ & $2.2{\times}10^{-3}$ \\
 & $32$ & $1.2{\times}10^{-3}$ & $7.5{\times}10^{-4}$ & $2.5{\times}10^{-3}$ & $\mathbf{2.2{\times}10^{-6}}$ & $5.1{\times}10^{-5}$ \\
 \bottomrule
 \end{tabular}
\end{table}

\paragraph{Bottleneck ablation.}
Table~\ref{tab:onedim_bneck} retrains the same spatial invertible map
with each bottleneck, keeping the training procedure fixed. Keeping the first
\(D\) entries of the flattened transformed field, the truncation used by nested-dropout flows, fails on every dataset. A fixed Fourier basis is
competitive only when the retained Fourier modes happen to suffice, and is up
to four orders of magnitude less accurate than OPINE.
Figure~\ref{fig:bneck_recon} shows a typical test profile of each dataset and
its reconstructions at \(D=4\).

\begin{table}[!ht]
 \centering
 \footnotesize
 \caption{Bottleneck ablation with the same spatial invertible map:
 median test NMSE.}
 \label{tab:onedim_bneck}
 \vspace{3pt}
 \begin{tabular}{lrccc}
 \toprule
 Dataset & $D$ & first $D$ entries & fixed Fourier basis & OPINE \\
 \midrule
Torus & $3$ & $3.3{\times}10^{-1}$ & $6.1{\times}10^{-3}$ & $\mathbf{6.0{\times}10^{-3}}$ \\
Torus & $4$ & $5.3{\times}10^{-1}$ & $7.1{\times}10^{-3}$ & $\mathbf{1.9{\times}10^{-5}}$ \\
Pulse & $2$ & $5.8{\times}10^{-1}$ & $6.1{\times}10^{-1}$ & $\mathbf{6.1{\times}10^{-4}}$ \\
Pulse & $4$ & $3.4{\times}10^{-1}$ & $4.1{\times}10^{-4}$ & $\mathbf{1.3{\times}10^{-4}}$ \\
Pulse width & $2$ & $4.5{\times}10^{-1}$ & $5.7{\times}10^{-1}$ & $\mathbf{1.1{\times}10^{-2}}$ \\
Pulse width & $4$ & $5.6{\times}10^{-1}$ & $9.6{\times}10^{-4}$ & $\mathbf{6.8{\times}10^{-4}}$ \\
Front & $2$ & $4.9{\times}10^{-1}$ & $5.2{\times}10^{-1}$ & $\mathbf{2.5{\times}10^{-5}}$ \\
Front & $4$ & $2.4{\times}10^{-1}$ & $3.9{\times}10^{-5}$ & $\mathbf{6.0{\times}10^{-6}}$ \\
 \bottomrule
 \end{tabular}
\end{table}

\paragraph{Subspace-update ablation.}
Table~\ref{tab:onedim_update} compares the update of Section~\ref{sec:training}
with a frozen subspace, weaker damping, exact recomputation of the POD
after every round, and differentiation through an eigendecomposition of the
transformed covariance at every step. Adapting the subspace matters on Pulse, where freezing it increases the error by a factor of about \(3\)
relative to the default and \(11\) relative to the best tracking variant,
and matters little on Torus, consistent with Appendix~\ref{app:pod}: the
damping changes the path of training rather than its target. Among the
tracking variants, the weaker damping \(\alpha=0.3\) is the most accurate,
so the default \(\alpha=0.03\) is conservative. Differentiating through an
eigendecomposition of the full transformed covariance at every step is the
most accurate option where it runs, but it is fragile and does not scale: with
a GPU eigensolver, all six runs failed because the solver did not converge on
the nearly repeated eigenvalues of the transformed covariance, and the results
shown were obtained with a CPU eigensolver at several times the cost per step.
Each step also requires the \(N\times N\) covariance of the entire training
set, which is impractical at the scale of the minimal channel
(\(N\approx2\times10^5\)). The tracking scheme avoids both problems.

\begin{table}[!ht]
 \centering
 \footnotesize
 \caption{Subspace-update ablation with the spatial invertible map: median test NMSE.
 The eigendecomposition variant was run with a CPU eigensolver; with a GPU
 eigensolver, all six runs failed.}
 \label{tab:onedim_update}
 \vspace{3pt}
 \begin{tabular}{lrccccc}
 \toprule
 Dataset & $D$ & frozen & $\alpha=0.03$ & $\alpha=0.3$ & exact & eigendecomposition \\
 \midrule
Torus & $3$ & $6.1{\times}10^{-3}$ & $6.0{\times}10^{-3}$ & $5.7{\times}10^{-3}$ & $5.8{\times}10^{-3}$ & $\mathbf{4.7{\times}10^{-3}}$ \\
Pulse & $2$ & $1.7{\times}10^{-3}$ & $6.1{\times}10^{-4}$ & $1.5{\times}10^{-4}$ & $6.2{\times}10^{-4}$ & $\mathbf{3.7{\times}10^{-5}}$ \\
 \bottomrule
 \end{tabular}
\end{table}

\paragraph{Size and cost of the coupling stack.}
Table~\ref{tab:onedim_eff} varies the number of blocks \(L\) and the
conditioner width \(w\). A compact spatial stack with \(L=4\) and \(w=32\)
comes within a factor of \(1.4\) of a stack with \(L=8\) and \(w=64\), the depth and width used for the
turbulent flows, with \(7\times\) fewer
parameters, and \(L=2\) remains within a factor of \(2\). Measured on a single
CPU thread under identical conditions for fields of \(N=128\) points, a
training step of the spatial stack costs \(13\), \(27\), and \(43\)\,ms for
\((L,w)=(2,32)\), \((4,32)\), and \((8,32)\), and one of the spectral stack
\(22\), \(49\), and \(103\)\,ms, about twice as much, because of its Fourier
transforms and per-wavenumber scale network.

\begin{table}[!ht]
 \centering
 \footnotesize
 \caption{Coupling-stack size: parameters and median test NMSE.}
 \label{tab:onedim_eff}
 \vspace{3pt}
 \begin{tabular}{llrrrc}
 \toprule
 Dataset & coupling & $L$ & $w$ & parameters & test NMSE \\
 \midrule
Torus ($D=3$) & spatial & $2$ & $32$ & $12{,}328$ & $7.4{\times}10^{-3}$ \\
Torus ($D=3$) & spatial & $4$ & $32$ & $24{,}656$ & $6.0{\times}10^{-3}$ \\
Torus ($D=3$) & spatial & $8$ & $64$ & $180{,}384$ & $4.4{\times}10^{-3}$ \\
Torus ($D=3$) & spectral & $2$ & $32$ & $19{,}524$ & $8.9{\times}10^{-3}$ \\
Torus ($D=3$) & spectral & $4$ & $32$ & $39{,}048$ & $6.3{\times}10^{-3}$ \\
KS ($D=8$) & spatial & $4$ & $32$ & $24{,}656$ & $3.8{\times}10^{-5}$ \\
KS ($D=8$) & spatial & $8$ & $64$ & $180{,}384$ & $3.2{\times}10^{-5}$ \\
KS ($D=8$) & spectral & $4$ & $32$ & $30{,}728$ & $8.9{\times}10^{-4}$ \\
 \bottomrule
 \end{tabular}
\end{table}

\paragraph{Nested prefixes.}
Table~\ref{tab:onedim_prefix} truncates a single model trained at \(K=6\) to
its ordered prefixes (Section~\ref{sec:training}). As on the turbulent flows
(Section~\ref{sec:nested}), the errors grow monotonically as the prefixes shorten and remain
below POD from \(D=3\) onward, but do not match models trained at each \(D\).

\begin{table}[!ht]
 \centering
 \footnotesize
 \caption{Ordered prefixes of one model trained at $K=6$ versus models
 trained at each $D$: median test NMSE (no model was trained at $D=5$). The two rows come from separate training runs.}
 \label{tab:onedim_prefix}
 \vspace{3pt}
 \begin{tabular}{lcccccc}
 \toprule
 & $D=1$ & $D=2$ & $D=3$ & $D=4$ & $D=5$ & $D=6$ \\
 \midrule
Torus, prefix & $3.8{\times}10^{-1}$ & $3.8{\times}10^{-2}$ & $1.9{\times}10^{-2}$ & $4.4{\times}10^{-4}$ & $2.3{\times}10^{-4}$ & $1.5{\times}10^{-5}$ \\
\quad trained at $D$ & $3.8{\times}10^{-1}$ & $1.3{\times}10^{-2}$ & $6.0{\times}10^{-3}$ & $1.9{\times}10^{-5}$ & -- & $1.5{\times}10^{-5}$ \\
Pulse width, prefix & $9.3{\times}10^{-1}$ & $4.9{\times}10^{-1}$ & $2.4{\times}10^{-1}$ & $1.8{\times}10^{-2}$ & $7.3{\times}10^{-3}$ & $2.1{\times}10^{-4}$ \\
\quad trained at $D$ & $5.2{\times}10^{-1}$ & $1.1{\times}10^{-2}$ & $1.9{\times}10^{-2}$ & $6.8{\times}10^{-4}$ & -- & $2.2{\times}10^{-4}$ \\
 \bottomrule
 \end{tabular}
\end{table}

\section{Experimental Details}
\label{app:details}

This appendix gives the training settings, the synthetic problem, the
additional ablations and results, the baselines, and the datasets of the
experiments outside the one-dimensional study.

\paragraph{OPINE training.}
Outside the one-dimensional study of Appendix~\ref{app:synthetic_results},
every OPINE model is trained at a single prescribed retained dimension. The
coupling stack uses \(L=8\) affine coupling blocks with conditioner width
\(w=64\) and is
optimized with AdamW \citep{loshchilov2019adamw} at learning rate \(10^{-3}\),
with a cosine-annealed schedule and gradient-norm clipping at \(1\). Training
runs for \(25\) rounds, each consisting of one epoch of stochastic
optimization followed by one damped
subspace update (Section~\ref{sec:training}) with \(\alpha=0.03\). The
log-scales of the coupling layers are bounded by \(2\tanh(\cdot/2)\). For the minimal channel, the basis is initialized with
\(12\) steps of orthogonal iteration (Eq.~\plaineqref{eq:power}); for
Kolmogorov flow, it is initialized from the exact POD basis of the training
data. The product \(C_zQ_K\) in Eq.~\plaineqref{eq:power} is accumulated over
minibatches of centered transformed states, so \(C_z\) is never formed; for
Kolmogorov flow it uses a random subset of \(20{,}000\) training states. The batch size is \(16\) for the minimal channel and \(64\) for Kolmogorov flow. Training is
performed in single precision, and invertibility is additionally checked in
double precision. We report the checkpoint with the lowest validation NMSE.
All turbulent-flow models use a single seed. OPINE has \(61{,}760\) and \(96{,}192\) parameters for Kolmogorov flow and the
minimal channel, and the CNN autoencoders have \(1.3\)--\(17.6\) and
\(2.5\)--\(19.1\) million parameters, increasing with \(D\).

\paragraph{Synthetic problem.}
For each of ten seeds, we sample \(32\) training, \(256\) validation, and
\(2000\) test phases. OPINE is trained for \(30\) rounds with learning rate
\(2\times10^{-3}\) and batch size \(256\), and the CNN autoencoder
(\(7\times10^4\) parameters) for \(400\) epochs. The test NMSE at \(D=2\) is
\(4.93\times10^{-3}\pm0.39\times10^{-3}\) for POD and
\(1.74\times10^{-4}\pm1.40\times10^{-4}\) for OPINE (mean and standard deviation over seeds); the ratio for individual seeds
ranges from \(11\) upward. The CNN autoencoder reaches a training NMSE of
\(1.25\times10^{-4}\), but its test NMSE ranges from \(3.9\times10^{-4}\) to
\(2.5\times10^{-1}\) across seeds, with median \(8.3\times10^{-4}\).

\paragraph{Nested prefixes of the turbulent-flow models.}
Table~\ref{tab:prefix_turb} compares the ordered prefixes of one model trained
at the largest \(D\) with models trained at each \(D\) and with POD
(Section~\ref{sec:nested}).

\begin{table}[!ht]
 \centering
 \footnotesize
 \caption{Test NMSE of the ordered prefixes of one OPINE model trained at
 the largest $D$, of OPINE models trained at each $D$, and of POD.}
 \label{tab:prefix_turb}
 \vspace{3pt}
 \setlength{\tabcolsep}{4pt}
 \begin{tabular}{llccccccc}
 \toprule
 & & $D=32$ & $64$ & $128$ & $256$ & $512$ & $1024$ & $2048$ \\
 \midrule
Kolmogorov & prefix & $0.357$ & $0.190$ & $0.0743$ & $0.0182$ & $0.00152$ & & \\
 & trained at $D$ & $0.209$ & $0.108$ & $0.0340$ & $0.00923$ & $0.00152$ & & \\
 & POD & $0.358$ & $0.195$ & $0.0804$ & $0.0261$ & $0.00610$ & & \\
\midrule
Minimal & prefix & $0.710$ & $0.634$ & $0.547$ & $0.460$ & $0.369$ & $0.280$ & $0.204$ \\
channel & trained at $D$ & $0.691$ & $0.608$ & $0.522$ & $0.440$ & $0.356$ & $0.277$ & $0.204$ \\
 & POD & $0.713$ & $0.641$ & $0.561$ & $0.480$ & $0.397$ & $0.314$ & $0.236$ \\
 \bottomrule
 \end{tabular}
\end{table}

\paragraph{Choice of the invertible map.}
Besides affine coupling layers, we tested invertible residual networks
\citep{behrmann2019iresnet,chen2019residual} and monotone elementwise maps
built from piecewise-quadratic and spline functions
\citep{huang2018neural,muller2019neural,durkan2019neural}. Invertible residual
networks were less expressive at a comparable number of parameters, because
the Lipschitz constraint that guarantees invertibility limits each block, and
their inverse requires a fixed-point iteration, which made the evaluation of
\(f_\theta^{-1}\) slow. Monotone maps needed deep stacks to couple distant grid
points, and these deep stacks suffered from vanishing gradients. Affine
coupling layers are inverted in closed form and gave the most accurate models
among the maps we tried.

\paragraph{CNN baseline training.}
The convolutional autoencoders are optimized with AdamW at learning rate
\(10^{-3}\) with a cosine-annealed schedule, and the reported checkpoint is
again selected by validation NMSE alone. The minimal channel uses \(50\) epochs with batch size \(32\), except that the
run at \(D=2048\) stopped early at epoch
\(30\) when its validation error stalled;
Kolmogorov flow uses \(30\) epochs with batch size \(64\). Padding follows the
boundary conditions of each dataset: circular in the spanwise direction and
zero in the wall-normal direction for the minimal channel, and circular in
both directions for Kolmogorov flow.

\paragraph{Datasets.}
The Kolmogorov dataset is generated with the pseudospectral JAX-CFD solver
\citep{kochkov2021machine} on the domain \((0,2\pi)^2\), using \(2/3\)
dealiasing and combined Crank--Nicolson/Runge--Kutta time stepping; we
verified that the trajectories are spectrally resolved and statistically
stationary after burn-in. After a burn-in of \(100\) time units, each of the
\(100\) trajectories provides \(1000\) snapshots at unit time intervals, and
the trajectories are split \(80/10/10\). The three-dimensional minimal channel spans
\(\pi h\times(\pi/2)h\times h\) in the streamwise, spanwise, and wall-normal
directions, with \(h\) the channel half-height, a no-slip floor, and a
stress-free top, on a \(32\times32\times64\) grid, and uses \(20{,}000\) snapshots of the three velocity components at
\(0.37\,h/u_\tau\) spacing with the chronological \(16{,}000/2{,}000/2{,}000\)
split of Section~\ref{sec:minimal3d}. The minimal-channel data are preprocessed as follows: for each velocity component and each wall-normal level, the
training-set mean profile is subtracted and the result is divided by the
training-set RMS profile.

\begin{figure}[!ht]
 \centering
 \includegraphics{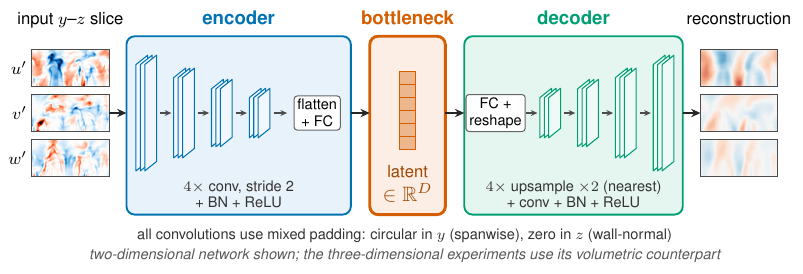}
 \caption{
 CNN autoencoder baseline, illustrated for a two-dimensional field with three velocity components: strided
 convolutions with boundary-aware padding compress the field to a
 $D$-dimensional latent vector, and a mirrored decoder reconstructs it. The
 three-dimensional experiments use its volumetric counterpart.
 }
 \label{fig:cnn_baseline}
\end{figure}

\end{document}